\documentclass[12pt,preprint]{aastex}

\shorttitle{Particle Acceleration in Sagittarius A*}
\begin{document}


\title{Stochastic Acceleration in the Galactic Center HESS Source}


\author{Siming Liu,\altaffilmark{1} Fulvio Melia,\altaffilmark{2,3} 
Vah\'e Petrosian,\altaffilmark{4} and Marco Fatuzzo\altaffilmark{5}}

\altaffiltext{1}{Los Alamos National Laboratory, Los Alamos, NM, 87545; liusm@lanl.edu}
\altaffiltext{2}{Physics Department and Steward Observatory, The University of Arizona, 
Tucson, AZ 85721; melia@physics.arizona.edu}
\altaffiltext{3}{Sir Thomas Lyle Fellow and Miegunyah Fellow.}
\altaffiltext{4}{Center for Space Science and Astrophysics, Department of Applied Physics, Stanford University, Stanford, CA 
94305; vahe@astronomy.stanford.edu}
\altaffiltext{5}{Physics Department, Xavier University, Cincinnati, OH 45207}


\begin{abstract}

Stochastic acceleration of charged particles interacting resonantly with a turbulent magnetic
field in a small accretion torus appears to be the likely mechanism responsible for much of
Sagittarius A*'s mm and shorter wavelength spectrum. The longer wavelength radiation is  
produced at larger radii by electrons either diffusing from smaller scales or accelerated
{\it in situ}. An important prediction of this model is the ejection of a significant flux of
relativistic protons from a magnetic-field-dominated acceleration site
into the wind-shocked medium surrounding the black hole. Recently, several air {\v
C}erenkov telescopes, notably HESS, have detected TeV emission from within $1^\prime$ of the
Galactic Center, with characteristics hinting at a pp-induced pion decay process for the
$\gamma$-ray emission. Given (i) that we now know the size of this acceleration region, where
Sagittarius A*'s 7-mm wavelength emission originates, and (ii) that we can now map the
wind-injected ISM within $\sim 3$ pc of the nucleus using the diffuse X-rays detected with
{\it Chandra}, it is feasible to test the idea that protons accelerated within $\sim 20$
Schwarzschild radii of the black hole produce the TeV emission farther out. We show that the
diffusion length of these particles away from their source guarantees a majority of TeV
protons scattering about once within $\sim 3$ pc of Sagittarius A*, and we demonstrate
that the proton power ($\sim 10^{37}$ ergs s$^{-1}$) produced in concert with the 7-mm 
radio emission matches the TeV luminosity well. The particle cascade generated by the
pp scatterings produces GeV $\gamma$-rays from $\pi^0$ decays, and bremsstrahlung,
inverse Compton, and synchrotron emission at longer wavelengths from secondary particles.
We compare these with current measurements and demonstrate that GLAST will detect this
source during its one-year all-sky survey.  This model explains why the TeV source is 
unresolved, yet does not vary on a time scale of a year or less, and it also accounts for 
the high-energy emission while retaining consistency with Sgr A*'s well-studied cm and mm 
characteristics.

\end{abstract}


%% subject header.  Objects should be in the appropriate "individual"
%% headers (e.g. quasars: individual, stars: individual, etc.) with the
%% additional provision that the total number of headers, including each
%% individual object, not exceed six.  The \objectname{} macro, and its
%% alias \object{}, is used to mark each object.  The macro takes the object
%% name as its primary argument.  This name will appear in the paper
%% and serve as the link's anchor in the electronic edition if the name
%% is recognized by the data centers.  The macro also takes an optional
%% argument in parentheses in cases where the data center identification
%% differs from what is to be printed in the paper.

%\object{NGC 6624}, \objectname[M 15]{NGC 7078},
%\object[Cl 1938-341]{Terzan 8})}


\section{Introduction}

Sagittarius A*, the compact radio source associated with the supermassive black hole (of mass
$M\sim 3.4\times 10^6\;M_\odot$; Sch\"odel et al. 2003, Ghez et al. 2003) at the Galactic 
Center, has a stratified emission structure in which the higher frequency emission is produced 
from progressively smaller volumes (see, e.g., Zhao, Bower, \& Goss 2001; Genzel et al. 2003; 
Baganoff et al. 2001; Melia, Jokipii, \& Narayanan 1992). The variable component of the 
near-IR (NIR) to X-ray emission fluctuates on a time scale of a few minutes (Porquet et al. 
2003; Ghez et al. 2004) and exhibits a quasi-periodic modulation during some long duration 
flares (Genzel et al. 2003; Belanger et al. 2005), hinting at an origin within an accretion 
torus with size equal to a few Schwarzschild radii ($r_S\equiv 2GM/c^2\simeq 10^{12}$ cm; Liu, 
Melia, \& Petrosian 2006). The spectral (Falcke et al. 1998) and polarization (Aitken et al. 
2000; Bower et al. 2005) characteristics of the mm/sub-mm radiation suggest that these too are 
produced within such a region (Melia, Liu, \& Coker 2000). The cm and longer wavelength emission, 
on the other hand, has to be produced on progressively larger scales to avoid being self-absorbed 
(Liu \& Melia 2001). Recent observations by Bower et al. (2004) show that indeed Sagittarius A*'s 
photosphere has a radius $\sim 24 r_S$ at 7 mm.

In earlier work (Liu, Petrosian, \& Melia 2004, hereafter LPM04), we showed that the 
stochastic acceleration (SA) of electrons interacting resonantly with plasma waves or 
turbulence generated via an MHD dissipation process (see, e.g., Miller, LaRosa, \& Moore 1996; 
Hamilton \& Petrosian 1992; Petrosian \& Liu 2004), can produce the nonthermal particles 
emitting much of Sagittarius A*'s spectrum (see, e.g., Zhao, Bower, \& Goss 2001; Genzel et 
al. 2003; Baganoff et al. 2001).  This mechanism, originally invoked to explain how magnetic 
reconnection can energize particles during solar flares, can also accelerate electrons in the 
magnetized Keplerian accretion torus within $\sim 5-10\,r_S$ of the black hole via a 
turbulence induced by the magneto-rotational instability (Balbus \& Hawley 1991; Melia, Liu, 
\& Coker 2001; Hawley \& Balbus 2002). The radiation emitted at the acceleration site can 
explain the mm/sub-mm and shorter wavelength observations of this source in both the 
quiescent and flaring states.
%and cannot be the dominant source of electrons producing cm wavelength emission.

An interesting prediction of this model is that, under most circumstances (see Fig. 1 in 
LPM04), the time required for energetic electrons to escape from the acceleration site is 
comparable to the acceleration and synchrotron cooling time scales so that there would be a 
significant outflux of accelerated particles.  The nonthermal ``halo'' or jet formed by the 
escaping electrons may contribute to the cm-wavelength emission farther out (see, e.g., Liu 
\& Melia 2001; LPM04; Zhao et al. 2004). In concert with the acceleration of electrons within
$\sim 5-10 r_S$ of the black hole, protons are also energized. But as we shall show in this
paper, their acceleration efficiency in this ``gas-dominated'' region (with Alfv\'{e}n 
velocity $v_A$ much less than the speed of light $c$) is not sufficient to produce an
observationally important flux away from Sagittarius A* itself. 

However, particles continue to be accelerated outside of the inner torus, as evidenced, e.g., by
the intensity and scale of the 7-mm emission region. We will therefore here also consider the 
{\it in situ} SA of particles in a larger ``magnetically'' dominated ($v_{\rm A}\simeq c$)
volume (extending out as far as $\sim 20-30r_S$), and demonstrate that the conditions required 
to account for Sagittarius A*'s 7-mm emission are in fact equally suited to the rapid acceleration 
of protons that escape and interact with the ambient medium surrounding Sagittarius~A* to produce 
$\gamma$-rays.

%On a bigger scale of tens of Schwarzschild radii, where the gas density drops and the medium 
%becomes magnetic-field dominated ($v_A>c$), protons as well as electrons can be accelerated 
%{\it in situ}. The electrons apparently produce the bulk of Sagittarius A*'s radio emission. 
%For example, Sagittarius A*'s size at 7 mm has recently been set observationally at $\sim 
%24\,r_S$ (Bower et al. 2004). 

Recently, the Galactic center (GC) has been identified by three air {\v C}erenkov 
telescopes at $\sim$ TeV energies: Whipple (Kosack et al. 2004), CANGAROO 
(Tsuchiya et al. 2004) and, most recently and significantly, HESS (Aharonian et al. 2004). 
The Hegra ACT instrument has also put a (weak) upper limit on GC emission at 4.5 TeV 
(Aharonian et al. 2002) and the Milagro water {\v C}erenkov extensive air-shower array
has released a preliminary finding of a detection at similar energies from the region 
defined as $l \in \{20^\circ, 100^\circ\}$ and $|b| < 5^\circ$ (Fleysher 2002).
We examine here the likelihood of these TeV photons being produced by the relativistic 
protons escaping the SA site near Sagittarius A*, and concentrate on the HESS results, 
though it should be noted that all of these observations lend crucial support to the notion 
that acceleration of particles to very high energies is taking place near (or at) the black hole. 
Observations pertinent to these high energy processes are described in \S\ \ref{obs}. In \S\ 
\ref{acs} we discuss the acceleration of electrons and protons by turbulence and the observational 
constraints on the magnetic field and gas density of an acceleration region roughly $20 r_S$ 
in size. The process of production of TeV photons is described in \S\ \ref{diff} and a summary 
is presented in \S\ \ref{dis}.

\section{Contextual Background}
\label{obs}

HESS detected a signal from the Galactic center in observations conducted over two
epochs (June/July 2003 and July/August 2003) with a $\sim6-9\sigma$ excess evident
over this period.  These data can be fitted by a power law with a spectral index
$2.21\pm0.09$, with a total flux above the instrument's 165 GeV threshold of 
$(1.82 \pm 0.22) \times 10^{-7}$ m$^{-2}$ s$^{-1}$ (there is also a 15-20\% error 
from the energy resolution uncertainty).

Intriguingly, the HESS data are difficult to reconcile with the EGRET GC source 
3EG~J1746-2851 (generally ascribed to pp-induced neutral pion decay at GeV energies; Fatuzzo 
\& Melia 2003).  In a re-analysis of select data from the 3EG catalog (Hartmann et al. 
1999), Hooper \& Dingus (2002) have shown that the GC is excluded at the $99.9\%$ confidence 
limit as the true position of 3EG~J1746-2851. The HESS source, on the other hand, is 
coincident within $\sim 1'$ of Sagittarius A*, though its centroid is displaced roughly 
$10^{\prime\prime}$ (corresponding to about $0.4$ pc at a distance $D=8$ kpc) to the East 
of the GC.  In addition, an extrapolation of the EGRET spectrum into HESS's range over-predicts 
(by a factor $\sim20$) the TeV $\gamma$-ray flux of the GC source.  Recently, Crocker et al. 
(2005), and Aharonian \& Neronov (2005a, 2005b), among others, have undertaken a detailed 
examination of the possible theoretical consequences of these new high-energy observations 
of the Galactic center, and inferred that we are apparently dealing with two separate sources.  
Perhaps both are associated with the SNR Sagittarius A East, but given its angular proximity 
to Sagittarius A*, the TeV source could be associated with the black hole itself.

What is not yet clear, however, is how the TeV $\gamma$-rays are produced. Models that invoke 
pp-induced neutral pion decays close to the black hole must contend with the collateral 
emission at other wavelengths (Markoff, Melia, \& Sarcevic 1999; Aharonian \& Neronov 
2005a) due to the secondary particles produced in the ensuing cascade. And because the inner 
$\sim 10\,r_S$-emitting region of Sagittarius A* has received extensive scrutiny over the 
years (see, e.g., Melia, 1992; Falcke et al. 1998; Bromley et al. 2001; Yuan, Quataert, \& 
Narayan 2003; Melia \& Falcke 2001; Melia 2006), non-hadronic models face similar difficulties in 
satisfying the constraints imposed across Sagittarius A*'s broadband spectrum---from radio 
waves to X-rays and $\gamma$-rays (but see Atoyan \& Dermer 2004 for a description of the 
``plerion" scenario, which avoids this complication). In the model we are proposing here, 
however, pp scattering events are quite rare in the acceleration site since
the ambient proton density is very low. Instead, most of the accelerated protons diffuse out 
of the system and scatter with the ISM surrounding the black hole.

Our understanding of the circum-black-hole environment at the GC has improved considerably in 
recent years, thanks to observations with {\it Chandra} and detailed hydrodynamic modeling of 
the diffuse X-ray emission it detected within $r\sim 10^{\prime\prime}$ of Sagittarius A*. 
The hot plasma producing this radiation, estimated at 
$\approx 7.6\times 10^{31}$ ergs s$^{-1}$ arcsec$^{-2}$ in the $2-10$ keV band, has an RMS 
electron density $\langle n\rangle\approx 26\,\eta_f^{-1/2}$ cm$^{-3}$ and a temperature 
$k_{\rm B}T\approx 1.3$ keV, with a total inferred mass of $\approx 
0.1\;M_\odot\,\eta_f^{1/2}$ (Baganoff et al. 2003), where $\eta_f$ is the volume filling 
factor.  We now understand that this hot gas is produced via collisions of winds of stars in 
the central cluster injected into the mini-cavity with a total mass rate of $\dot{M}\simeq 
3\times 10^{-3} M_\odot$ yr$^{-1}$ (Rockefeller et al. 2004).  In the most recent census, 
some 25 bright, young wind sources have been identified and their mutual interactions account 
for the entire X-ray flux observed with {\it Chandra}. For a typical wind velocity $v=1000$ 
km s$^{-1}$, the average (pre-shock) gas density in the central $\sim 1$ pc region is 
$\langle n_0\rangle\sim 200$ cm$^{-3}$. We therefore have a good handle on the physical 
conditions in this region.  In addition, this medium appears to be threaded by a rather 
strong magnetic field $\sim 0.1-1$ mG (Yusef-Zadeh et al. 1996), through which the escaping 
relativistic protons must diffuse. Note that this measured magnetic field is roughly in equipartition 
with the gas.

\section{The Acceleration Site}
\label{acs}

To date, the most reliable measurement of Sagittarius A*'s intrinsic size has been made with
closure amplitude imaging using the VLBA (Bower et al. 2004). At a wavelength of 7 mm, the 
source is confined to the inner $\sim 24$ Schwarzschild radii, though the size is variable
on a time scale of hours to weeks and sometimes reaches $\sim 60\,r_S$.  Correspondingly, 
Sagittarius A*'s average flux density at this wavelength is $\approx 1.5$ Jy (e.g., Falcke et 
al. 1998). These set strict constraints on the SA model, because the acceleration must not 
only be confined to a region of known size ($\sim 24\;r_S$), but must also produce a 
particle distribution that accounts for the measured radiative output at this wavelength. 

A complete treatment of SA by waves requires a solution of the coupled kinetic equations for the   
pitch-angle-averaged particle distribution $N(E, t)$ and the wave spectrum ${\cal W}({\bf k})$:
\begin{eqnarray}
\label{HOMOG}
   {\partial N \over \partial t}
& = & {\partial \over \partial E}\left[D_{EE}{\partial N \over \partial E}
 - (A-\dot E_L) N\right]
 - {N \over T_{\rm esc}} +{\dot Q}\,, \\
   {\partial {\cal W}({\bf k}, t) \over \partial t}
& = &\dot{Q}_{\cal W}({\bf k}, t) - \Gamma({\bf k}){\cal W}({\bf k}, t) +{\partial
\over\partial k_i}\left[D_{ij}{\partial\over \partial k_j}{\cal W}({\bf k}, t)\right] - 
{{\cal W}({\bf k}, t) \over T^{\cal W}_{\rm esc}({\bf k})}\,,
\label{WHOMOG}
\end{eqnarray}
where $E=\gamma-1$, $\gamma$ and ${\bf k}$ are, respectively, the particle kinetic energy in 
units of the rest mass energy, Lorentz factor and the wave vector. The loss rate, the source 
and escape terms for the particles are given by ${\dot E}_L$, ${\dot Q}$ and $N(E)/T_{\rm 
esc}$, respectively. The terms on the right-hand-side of Equation (\ref{WHOMOG}) represent 
the wave generation, damping, cascade, and leakage processes. The energy diffusion coefficient 
$D_{EE}(E)$ and the function $A(E)=(D_{EE}/E)(2-\gamma^{-2})/(1+\gamma^{-1})$ depend on ${\cal 
W}({\bf k})$, and the wave damping rate $\Gamma({\bf k})$ depends on $N(E)$. That is how the 
equations are coupled.

SA of particles is dominated by the transit-time damping and cyclotron-resonance processes.
%The former only increases the parallel component of the particle momentum (Miller et al. 1996).
%In the absence of an efficient scattering process, the transit-time damping acceleration can be 
%suppressed significantly at relativistic energies. 
%with the latter prevail in strongly magnetized plasmas. 
Recent progress in studying the scattering and acceleration of charged particles by MHD turbulence 
indicates that fast mode waves play the dominant role (Yan \& Lazarian 2002). Fast wave turbulence 
is isotropic and has a power-law spectrum with an index of $q=1.5$ in the inertial range (Cho, 
Lazarian \& Vishniac 2002). At small scales, the waves are subjected to collisionless damping, 
which depends on the wave propagation angle with respect to the magnetic field $\theta$ 
(Stepanov 1958; Ginzburg 1961; Petrosian, Yan, \& Lazarian 2005):
\begin{equation}
\Gamma({\bf k}) \simeq u(\pi\beta_p\delta)^{1/2}{kv_{\rm A} \sin^2{\theta}\over 2\cos{\theta}}\;,
\ \ \ \  {\rm for}\ \ \ \  1\gg\beta_p\gg\delta\;,
\end{equation}
where $u$, $\delta=m_e/m_p$, and $\beta_p=4\pi P_{\rm gas}/B^2$ are the ratio of the proton to 
electron energy density, the electron to proton mass, and the gas to magnetic field pressure, 
respectively. The wave spectrum will cut off when this 
damping rate dominates the cascade rate $\tau^{-1}_{\rm cas}(k) \simeq v_{\rm A}f_{\rm 
turb}(k_{\rm min}k/2\pi)^{1/2}$, where $2\pi/k_{\rm min}$ gives the turbulence injection length scale 
and $f_{\rm turb}=[8\pi\int{\cal W}(k){\rm d}k]/B^2<1$ is the ratio of the turbulence to magnetic 
field energy density. Most of the waves at small scales therefore propagate along the magnetic field 
line with the propagation angle $\theta<\theta_{\rm cr}(k)=(2f^2_{\rm turb}k_{\rm min}/\pi^2u^2
\beta_p\delta k)^{1/4}$. To study particle acceleration by such turbulence, one may represent 
the latter as parallel propagating waves with spectrum
${\cal W} \propto k^{-3/2}\theta_{\rm cr}^2\propto k^{-2}$. 

Almost all of the observed emission from the direction of Sagittarius A* is produced by nonthermal 
relativistic particles ($E\simeq \gamma\gg 1$ and $A\simeq 2D_{EE}/E$). As such, the characteristic 
interaction time $\tau_p$, the pitch-angle-averaged scattering ($\tau_{\rm sc} \equiv \langle 
[1-\mu^2]^2/D_{\mu\mu}\rangle$) and acceleration ($\tau_{\rm ac}\equiv E^2/D_{EE}$) times, 
and the escape time, can all be expressed in units of $\tau_{\rm tr}\equiv R/c$, the light 
transit time for an acceleration region of size $R$ (Dung \& Petrosian 1994; Petrosian \& Liu 
2004):
\begin{eqnarray}
\tau_p &=& {4\over \pi (q-1)f_{\rm turb}\Omega}\left({ck_{\rm min}\over 
\Omega}\right)^{1-q}\simeq1.3{\tau_{tr}\over (q-1)f_{\rm turb}Rk_{\rm min}}\left({ck_{\rm 
min}\over \Omega}\right)^{2-q}
%\left({\Omega_e\over \Omega}\right)^q
\,, \\
\tau_{sc} 
&\simeq& {\tau_p} \gamma^{2-q}
\simeq 1.3{\tau_{tr}\over (q-1)f_{\rm turb}Rk_{\rm min}}\left({\gamma 
ck_{\rm min}\over \Omega}\right)^{2-q}\,,  \label{sc} \\
\tau_{\rm ac} 
&\simeq& {2(\beta_{\rm A}^2+1)\tau_p\over \beta_{\rm A}^{2}} \gamma^{2-q}
\simeq {2.6(\beta_{\rm A}^2+1)\over \beta_{\rm A}^2}{\tau_{tr}\over (q-1)f_{\rm 
turb}Rk_{\rm 
min}}\left({\gamma ck_{\rm min}\over 
\Omega}\right)^{2-q}\,, \label{acc} \\
T_{\rm esc} &=&{\tau_{tr}^2\over \tau_{sc}} +2^{1/2}\tau_{\rm tr}\simeq 
\left[0.8(q-1)f_{\rm turb}Rk_{\rm 
min}\left({\gamma ck_{\rm min}\over \Omega}\right)^{q-2} + 1.4\right]\tau_{\rm tr}\,,
\label{tesc}
\end{eqnarray}
where $\mu$, $D_{\mu\mu}$, and $\Omega$ are, respectively, the cosine of the particle pitch angle 
with respect to the large scale magnetic field $B$, the pitch angle scattering rate, and the 
nonrelativistic gyrofrequency of the particles, and $2\pi/k_{\rm min}\lesssim R$. The 
(dimensionless) Alfv\'{e}n velocity is defined as 
\begin{equation}
\beta_{\rm A} \equiv {B\over (4\pi nm_pc^2)^{1/2}} =
7.3 \left({B\over 1 {\rm G}}\right)\left({n\over 1{\rm cm}^{-3}}\right)^{-1/2}
\simeq
{v_{\rm A}/c\over [1-(v_{\rm A}/c)^2]^{1/2}}\,.
\label{alf}
\end{equation} 
Note that, for $\beta_{\rm A}\ge 1$, the phase velocity of the waves is $v_{\rm A}/c=w/(ck)
\simeq\beta_{\rm A}/(1+\beta_{\rm A}^2)^{1/2}$.

For $q=2$, the acceleration and escape time scales become energy independent. When the loss 
processes are unimportant (i.e., $\dot{E}_L\ll A$), Equation (\ref{HOMOG}) can be reduced to
\begin{equation}
   {\partial N \over \partial t}
= {\partial \over \tau_{\rm ac}\partial E}
\left[E^2{\partial N \over \partial E} - 2 E N\right] 
 - {N \over T_{\rm esc}} +{\dot Q}\,, \\
\label{q2}
\end{equation}
whose steady-state solution is $N\propto E^{-p}$ above the particle injection energy, where 
the spectral index $p$ is determined by the ratio of $\tau_{\rm ac}$ to $T_{\rm esc}$:
\begin{equation}
p = \left({9\over 4} +{\tau_{\rm ac}\over T_{\rm esc}}\right)^{1/2}-0.5>1.0\,.
\label{index}
\end{equation}

There are four primary parameters in the SA model (LPM04): $R$, $f_{\rm turb}Rk_{\rm min}$, $n$ 
and $B$. For 
Sagittarius A*, $R\simeq 20 r_S$ at the 
wavelength of interest (i.e., 7 mm). The observed $\gamma$-ray spectral index and the 
radio to IR spectrum set strict constraints on the other three. 
To produce a power-law spectrum of relativistic protons with $p=2.2$, which gives rise to a 
$\gamma$-ray spectral index of the same value via pp scattering processes, Equations 
(\ref{index}) and (\ref{tesc}) show that $\tau_{\rm ac}\simeq 5.0\,T_{\rm esc}= 
(4f_{\rm turb}Rk_{\rm min}+7)\tau_{\rm tr}$. Combining this with Equation (\ref{acc}), one has
\begin{equation}
\beta_{\rm A} = \left({2.6\over 4[f_{\rm turb}Rk_{\rm min}]^2+7f_{\rm turb}Rk_{\rm min} 
-2.6}\right)^{1/2}\,.
\label{betaf}
\end{equation}
%where we have used $Rk_{\rm min}=2\pi$. 
Since $f_{\rm turb}\le1$, this sets a constraint on $\beta_{\rm A}$.
%plasma has to be 
%strongly magnetized with $\beta_{\rm A}>0.11$.
%More accurate numerical calculations of the 
%acceleration and scattering times show that $\beta_{\rm A} >0.15$. 
%In Figure \ref{fig2}, we demonstrate this constraint by excluding the upper-left corner 
%of the $n-B$ plane.

%Next, we consider the case $\beta_{\rm A}\ge1$. Because $T_{\rm 
%esc}>1.4\tau_{\rm tr}$, $\tau_{\rm ac} \simeq \tau_{\rm sc}$ must be larger than 
%$7.1\,\tau_{\rm tr}$, which implies $f_{\rm turb}\le 0.028$. Therefore $\tau_{\rm sc}$ is 
%much larger than $\tau_{\rm tr}$, and we can ignore the diffusion term in the expression for 
%the escape time: $T_{\rm esc} = 1.4\,\tau_{\rm tr} = 940\,(R/20r_S)$ s and $\tau_{\rm ac} = 
%7.1\,\tau_{\rm tr} = 4.7\times10^3 (R/20\,r_{S})\ {\rm s}\,.$ Then $f_{\rm turb}\simeq 
%0.02$, though its exact value depends on $\beta_{\rm A}$ because of the weak dependence of 
%$\tau_{\rm ac}$ and $\tau_{\rm ac}$ on $\beta_{\rm A}$.

Next we consider the constraints imposed on $B$ and $n$ by the 7 mm observations. The turbulence 
discussed above accelerates both protons and electrons to a power-law distribution with $p=2.2$ 
in the relativistic energy range. The source of electrons for this process is likely the hot flow 
accreting toward the black hole. 
%The temperature of this plasma is typically $\sim 10^{10}-10^{11}$ K 
%(see, e.g., Liu \& Melia 2001). 
We therefore assume that the injected electrons have a mean Lorentz factor $\gamma_{\rm in}=10$.
%, corresponding to a temperature of $6\times 10^{10}$ K. 
So the electron distribution has a rising spectrum below this energy and a power-law 
spectrum above it (Park \& Petrosian 1995). The distribution cuts off when the synchrotron 
cooling time 
\begin{equation}
\tau_{\rm syn} = {9m_e^3c^5\over 4e^4B^2\gamma} = 78 \left({B\over 100 {\rm 
G}}\right)^{-2}\left({\gamma\over 10^3}\right)^{-1}{\rm \ s}
\end{equation} 
equals the acceleration time $\tau_{\rm ac}= 5.0T_{\rm esc} = 670(4f_{\rm turb}Rk_{\rm min}+7)
(R/20r_S)$ s. Here, $m_e$ is the electron mass and $e$ is its charge. One therefore 
gets a high-energy cutoff 
\begin{equation}
\gamma_{M}= {17\over1+0.6f_{\rm turb}Rk_{\rm min}}\left({B\over 100 {\rm 
G}}\right)^{-2}\left({R\over20r_S}\right)^{-1}\,.
\end{equation}
%which should be larger than $\gamma_{\rm in}$. This gives 
%\begin{equation}
%B<
%130\,\left({1\over1+3.6f_{\rm turb}}\right)^{1/2}
%\left({R\over20\,r_S}\right)^{-1/2}
%\left({\gamma_{\rm in}\over10}\right)^{-1/2}\,{\rm G}\,
%\le130\,
%\left({R\over20\,r_S}\right)^{-1/2}
%\left({\gamma_{\rm in}\over 10}\right)^{-1/2}
%\,{\rm G}\,.
%\end{equation}
%Using equations (\ref{betaf}) and (\ref{alf}), we have
%\begin{equation}
%n \le 2.5\left({B\over 0.1{\rm G}}\right)^2\left\{\left[
%\left({B\over 130{\rm G}}\right)^{-2}
%\left({R\over20\,r_S}\right)^{-1}
%\left({\gamma_{\rm in}\over10}\right)^{-1}
%-0.5\right]^2-0.45\right\}{\rm cm}^{-3}\,.
%\end{equation}
The corresponding frequency of the synchrotron emission is 
\begin{equation}
\nu_M = 
{3eB\gamma_M^2\over 4\pi m_ec}=120\,\left({1\over1+0.6f_{\rm turb}Rk_{\rm min}}\right)^{2}
\left({B\over100\ {\rm 
G}}\right)^{-3} 
\left({R\over 20\,r_S}\right)^{-2}{\rm GHz}\;,
\end{equation}
which should be higher than the observed frequency $43$ GHz.  
We choose $\nu_M\ge 60$ GHz (which is $40\%$ above the observed value), giving
\begin{equation}
B\le 
126\,\left({R\over20\,r_S}\right)^{-2/3}
%\left({\gamma_{\rm in}\over10}\right)^{1.2}
\left({1\over 1+0.6f_{\rm turb}Rk_{\rm min}}\right)^{2/3}\, {\rm G}
< 126\,\left({R\over20\,r_S}\right)^{-2/3}
%\ge
%4\,
%\left({R\over20\,r_S}\right)^{-2/3}
\, {\rm G}\,.
\end{equation} 
%From equations (\ref{betaf}) and (\ref{alf}), we have
%\begin{equation}
%n \ge 2.5\left({B\over 0.1{\rm G}}\right)^2\left\{\left[
%\left({B\over 11\ {\rm G}}\right)^{-3/2}
%\left({R\over20\,r_S}\right)^{-1}
%-0.5\right]^2-0.45\right\}{\rm cm}^{-3}\,.
%\end{equation}
This constraint is indicated by the vertical line in Figure~1.

\begin{figure}[htb]
\begin{center}
\includegraphics[height=12cm]{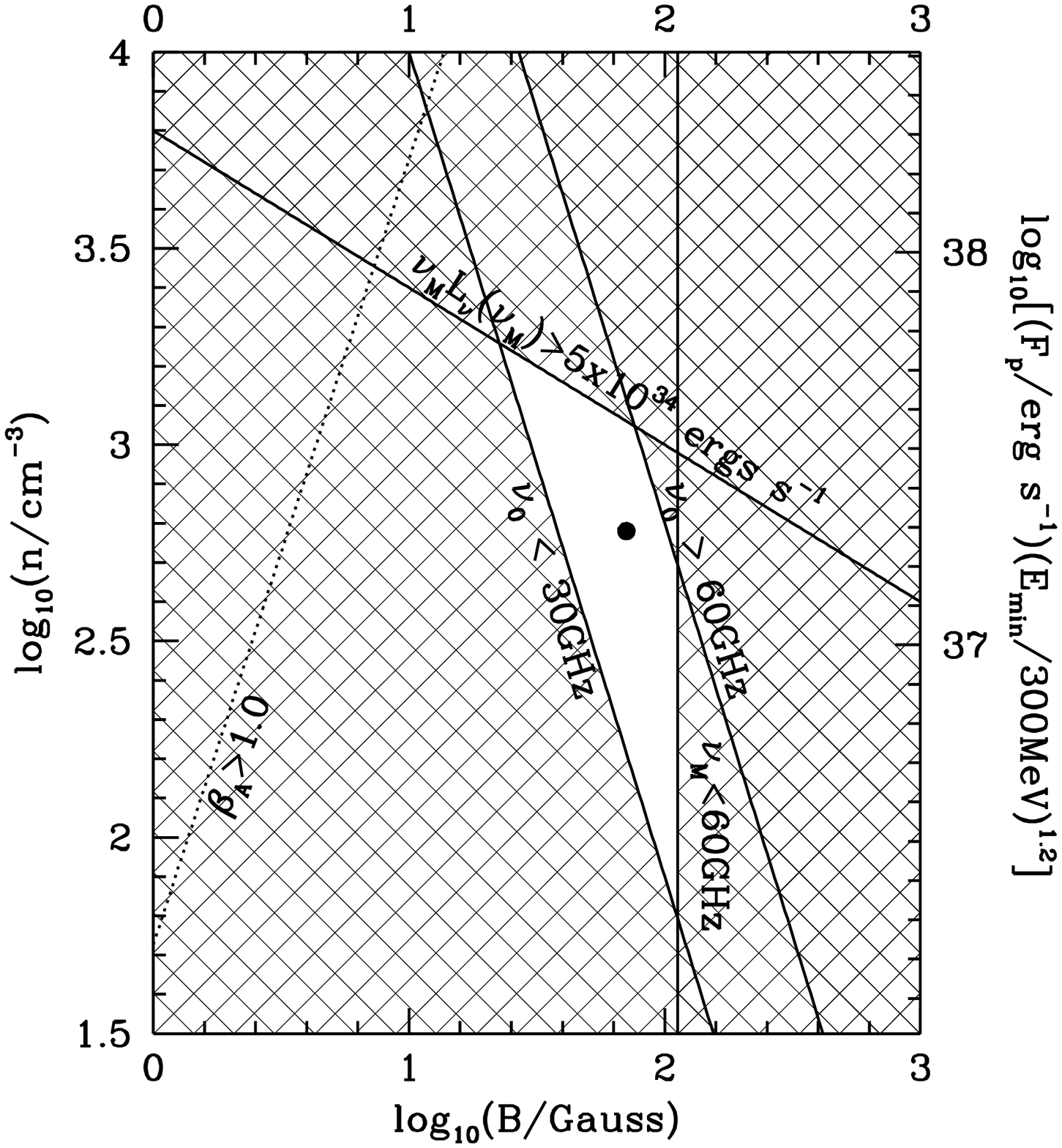}
\end{center}
\caption{Constraints on the density and magnetic field of the acceleration site, where electrons 
produce the observed 7 mm emission while protons escape toward large radii and produce the HESS 
$\gamma$-ray signal via collisions with the ambient protons. Here we have assumed that $\gamma_{\rm 
in} = 10$ and $R=20\, r_S$. The dashed line corresponds to $\beta_{\rm A}=1$.
%Parameter space of the SA model for Sagittarius A*. 
At 7 mm the source is optically thin (thick) in the lower-left (upper-right) portion of the 
figure. The constraint $B\lesssim120$ G comes from the requirement that the synchrotron emission 
cuts 
off above $60$ GHz. The upper area is excluded because the total synchrotron power from 
the acceleration site can not exceed the radio to IR luminosity of Sagittarius A*.
Depending on the details of the source geometry, the 7 mm observations may be explained with $B$ 
and $n$ in the blank (central) region of this figure. For example, the black dot corresponds to an 
illustrative uniform sphere model.  The right axis indicates the power in escaping protons with 
energy $E>0.1$ TeV.  
\label{fig2} 
}
\end{figure}

Because Sagittarius A* has a flat radio spectrum with $\alpha\simeq -0.3$ ($F_\nu\propto 
\nu^{-\alpha}$), the optical depth of this nonthermal source at 7 mm $\tau_0$ has to be $\sim 
1$ to avoid producing emission in excess of the observed flux below (for $\tau_0<1$) or above 
(for $\tau_0>1$) 7 mm. We require $30{\rm GHz}\le\nu_0\le 60{\rm GHz}$, where $\tau_0(\nu_0)=1$ 
for a source extent of $R=20r_S$, i.e. (Pacholczyk 1970)
\begin{equation}
0.08\le
\left({n\over 10^3{\rm cm}^{-3}}\right)
\left({\gamma_{\rm in}\over10}\right)^{1.2}
\left({B\over 100{\rm G}}\right)^{2.1}
\left({R\over20\,r_S}\right)
\le
0.68\,,
\end{equation}
which excludes the lower-left and upper-right portions of the $B-n$ plane (Fig. \ref{fig2}). 
The luminosity associated with this emission (assuming a spherical source) would be given by 
(Pacholczyk 1970)
\begin{eqnarray}
\nu_ML_\nu(\nu_M) &=&  4.9\times 10^{34} \\ 
&&\left({n\over 10^3{\rm cm}^{-3}}\right) 
\left({\gamma_{\rm in}\over 10}\right)^{1.2}
\left({B\over 100 {\rm G}}\right)^{0.4} 
%\left({3V\over 4\pi R^3}\right)
\left({R\over 20\,r_S}\right)^{2.2} 
\left({1\over 1+0.6f_{\rm turb}Rk_{\rm min}}\right)^{0.8}
{\rm ergs\ s}^{-1}\,,\nonumber
\end{eqnarray}
which must be lower than the radio to NIR luminosity of $\sim5\times 10^{34}$ ergs s$^{-1}$. In 
combination with Equation (\ref{betaf}), this constraint excludes the upper portion of the $B-n$ 
plane (Fig. \ref{fig2}). Therefore, to reconcile the radio to IR and $\gamma$-ray emissions with 
the SA model, the plasma must be strongly magnetized 
with $\beta_{\rm A}>1$. The blank region near the middle of Figure \ref{fig2} shows the allowed 
parameter space, when all these constraints are taken into account. The exact properties of the 
plasma required to explain the 7 mm observations depend on the source structure and geometry. As an 
illustration, we consider a uniform spherical model with radius $R=20\,r_S$ (roughly consistent with 
the VLBA measurement). Calculating the accelerated electron spectrum and consequent radio emission 
yields a best fit model for the 7 mm data (with $B=70$ G, $n=600$ cm$^{-3}$, and $f_{\rm 
turb}Rk_{\rm min} = 0.016$) corresponding to the black dot near the middle of Figure \ref{fig2}.

\begin{figure}[htb]
\begin{center}
\includegraphics[height=12cm]{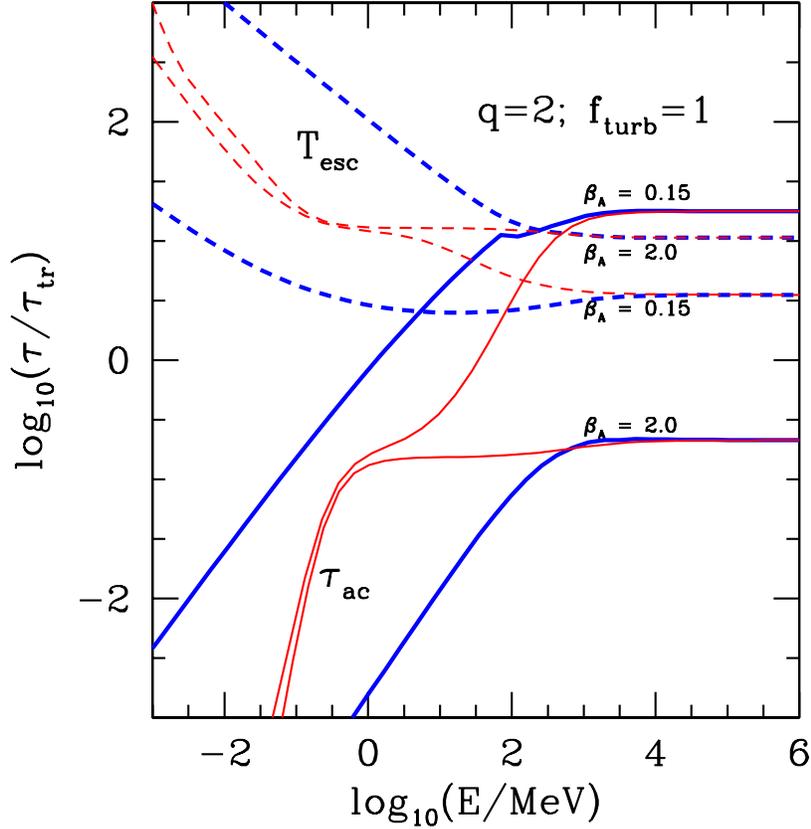}
\end{center}
\caption{
The acceleration (solid) and escape (dashed) time scales for electrons (thin) 
and protons (thick), in plasmas with $\beta_{\rm A} = 0.15$ and $2.0$ (indicated in the figure). 
The time scales are shown in units of the light transit time $\tau_{\rm tr} = R/c$. The turbulence 
is assumed to be in energy equipartition with the magnetic field and $q=2$. In a weakly magnetized 
plasma ($\beta_{\rm A}<1.0$), whistler waves can accelerate low-energy electrons to a Lorentz factor 
$\beta_{\rm A}m_p/m_e$,
%where $m_p/m_e$ is the proton to electron mass ratio, 
causing a sharp rise 
in the electron acceleration time with energy beyond 1 MeV. The acceleration of nonrelativistic 
electrons is dominated by high frequency electromagnetic waves. In a strongly magnetized plasma 
($\beta_{\rm A}>1.0$), the whistler wave branch vanishes, the acceleration time of 
relativistic electrons is almost energy-independent. 
\label{fig1}
} 
\end{figure}

Now, in concordance with the acceleration of electrons to produce Sagittarius A*'s 7 mm 
emission, protons are also energized.  Figure \ref{fig1} shows the 
acceleration and escape time scales for electrons ({\it thin}) and protons ({\it thick}) in 
plasmas with $\beta_{\rm A}=0.15$ and $2.0$.  The time scales are given in 
units of $\tau_{\rm tr}$, and we assume $f_{\rm turb} Rk_{\rm min} = 2\pi$ for this 
calculation. In the general case, $\tau_{\rm ac}\propto \tau_{\rm sc}\propto f_{\rm 
turb}^{-1}\ge1$. From these time scales, we see that a hard high energy proton spectrum can only 
be produced in plasmas with $\beta_{\rm A}\ge0.15$ and  
low-energy protons have an escape time much longer than their acceleration time. Thus, most 
of these particles are energized efficiently to about $E_{\rm min}\sim300$ MeV, above which 
the acceleration and escape times become comparable and the protons attain a power-law 
distribution. For such a spectrum, we have $N(E) = 
(p-1)\,n\,E_{\rm min}^{p-1}\,E^{-p}$, where we have used the fact that the high cutoff energy 
$E_{\rm max}\simeq 2\pi eB/k_{\rm min} = 4.2\times 10^{17}$ eV is much larger than the 
`turnover' energy $E_{\rm min}$. The ensuing energy flux associated with protons with $E\ge 
E_o=0.1$ TeV is given by
\begin{eqnarray}
&F_p& = {p-1\over p-2}{VnE_{\rm min}^{p-1} E_o^{2-p}\over T_{\rm esc}} \\
&=& 3.2\times 
10^{37}\,\left({n\over 10^3{\rm cm}^{-3}}\right)\left({E_{\rm
min}\over 300 {\rm MeV}}\right)^{1.2}\left({E_o\over 0.1{\rm TeV}}\right)^{-0.2}
\left({3V\over 4\pi R^3}\right)\left({R\over 20\,r_S}\right)^2 {\rm ergs\ s}^{-1}\,. 
\nonumber
\end{eqnarray}
With all other parameters being fixed, $F_p$ only depends on $n$, and is shown on the 
right axis of Figure \ref{fig2}.  Note that the illustrative uniform spherical model for the 
7 mm observations of Sagittarius A* (introduced above) corresponds to a proton energy flux $F_p 
\approx 1.9 \times 10^{37}$ ergs s$^{-1}$. The total power associated with the proton flux is 
$\sim 6.0\times 10^{37}$ erg s$^{-1}$. The power injected into the fast wave turbulence can be 
estimated as $\sim (4\pi/3)R^2v_{\rm A}B^2f^2_{\rm turb}\sim 6.0\times 10^{37}$ergs s$^{-1}$. 
Most of the dissipated turbulence energy then goes into the protons. And for the model parameters 
$\theta_{\rm cr}<0.5$ for $k>k_{\rm min}$ (note that $u\sim E_{\rm min}/\gamma_{\rm in}m_e 
c^2\simeq 60$), the approximation of the turbulence as parallel propagating waves is therefore 
appropriate.

\section{The PP Interaction Region}
\label{diff}

Once the relativistic protons leave the acceleration site, they diffuse through the medium 
surrounding Sagittarius A*, as described in \S\ 2 above. The transport of cosmic rays through 
chaotic magnetic fields has been considered by several authors, among them Giacalone and 
Jokipii (1999), and Casse, Lemoine, and Pelletier (2001).  These particles effectively 
execute a random walk through the ISM with a mean free path $c\tau_{\rm sc}$. There is no 
evidence of a large scale magnetic field on the extension corresponding to the HESS 
point-spread-function. In addition, no structure has been detected within the mini-cavity on angular 
scales referenced to {\it Chandra}'s spatial resolution of $\sim 1$ arcsec. Even more constraining 
are the numerical simulations for the diffuse X-ray emission, which show that shocks resulting from 
wind-wind collisions are typically ten times smaller than the separation between stars. This 
corresponds roughly to a coherence scale $\sim 10^{16}$ cm.  We therefore infer that $f_{\rm 
turb}\simeq1$ and $k_{\rm min}\simeq 6\times 10^{-16}$ cm$^{-1}$. Because the temperature of the 
ambient gas is low ($\beta_p\sim 1$), the damping of fast mode waves interacting with relativistic 
protons is unimportant. The turbulence is isotropic with $q=3/2$ up to $k_{\rm cr}$, which can be 
estimated with $\theta_{\rm cr}(k_{\rm cr})=1$. We therefore have $k_{\rm cr} \simeq 400 k_{\rm 
min}/\beta_p$. 

Combined with Equation (\ref{sc}), this therefore yields a scattering time in this region surrounding
Sagittarius A* of $\tau_{\rm sc}\simeq 6.4\times 10^4 (E/10\,{\rm TeV})^{1/2}(B/0.1\,{\rm 
mG})^{-1/2}$ s. The HESS source may be as large as $R= 3$ pc (Aharonian et al. 2004; Aharonian \& 
Neronov 2005b). Thus, the total path length of a relativistic proton in the medium surrounding 
Sagittarius A* is $l_{\rm tot}= R^2/(\tau_{\rm sc} c) \simeq 4.5\times10^{22}(E/10\, {\rm 
TeV})^{-1/2}(B/0.1\,{\rm mG})^{1/2}$ cm. On the other hand, the cross section $\sigma_{pp}$ 
for pp scattering is weakly dependent on energy, and may be taken as $\sim 40$ mb in the 
energy range of interest (see, e.g., Crocker et al. 2004).  Therefore the mean free path for 
a pp scattering event is $\sim [\langle n_0\rangle\sigma_{pp}]^{-1}\approx 1.3\times 10^{23} [\langle 
n_0\rangle/200\, {\rm cm^{-3}}]^{-1}$ cm, so $l_{\rm tot}\langle n_0\rangle\sigma_{pp}\sim O(1)$.  In 
other words, most of the escaping relativistic protons scatter with background protons about once 
in the wind-shocked medium surrounding Sagittarius A*. 
Because the corresponding pp scattering time is $\sim 10^5$ yrs, which is longer than the time 
required for the wind-injected ISM to reach equilibrium (Rockefeller et al. 2004), any variability 
in the TeV emission would be associated primarily with density fluctuations on large spatial scales, 
rather than changes in the source. We would expect the source to be stable on a time scale of 
a year or less.\footnote{Equation (\ref{acc}) shows that the corresponding acceleration time 
in this region is $\tau_{\rm ac} \simeq 7.5\times 10^8 \tau_{\rm sc} 
(\langle n_0\rangle/200\,{\rm cm}^{-3}) (B/0.1\,{\rm mG})^{-2}$, which is longer than any other 
relevant time scales, so the acceleration of protons by the turbulence outside of Sagittarius A*
itself may be ignored.}

The particle cascade generated by scatterings between the relativistic protons diffusing
away from the acceleration site and protons in the ambient medium produces additional  
spectral components due to neutral pion decays and emission by secondary leptons.
Neutral pion decays produce a $\gamma$-ray spectrum that dominates above $\sim 100$ MeV.
At lower energies, most of the radiative emission is due to the decay products of charged
pions. These leptons radiate via synchrotron, bremsstrahlung, and inverse Compton scattering
with the strong UV and IR fields bathing the inner few parsecs of the Galaxy (Telesco,
Davidson \& Werner 1996; Davidson et al. 1992;  Becklin, Gatley \& Werner 1982).  A detailed 
description of how the total emission from these various processes can be calculated appears
in Fatuzzo \& Melia (2003). For a proton energy flux $F_p \approx 1.6\times 10^{37}$ ergs 
s$^{-1}$ (above 300 MeV with $p=2.25$), the broadband spectrum resulting from these processes is 
shown in Figure~3. For this calculation, we assume that the protons are scattered only once and 
the magnetic field strength in the pp interaction zone was set at $1.0$ mG. The currently 
available data (or upper limits) are also shown here for comparison with our model. We also 
include the GLAST single energy sensitivity for a one-year all-sky survey, which shows that the 
HESS $\sim 100$ MeV--$100$ GeV counterpart may also be detectable. Note that since cooling time 
of TeV electrons and positrons in the pp interaction region is considerably shorter than the pp 
scattering time ($\sim 4\times 10^{12}$ s), their distributions reach a cooling broken-power 
spectrum in the steady-state.

%The integrated power in protons (above $0.1$ TeV) required to account for the HESS spectrum 
%is $\sim 3\times 10^{36}$ ergs s$^{-1}$, taking into account the fractional energy that 
%actually ends up in the $\pi^0$ channel (see Crocker et al. 2004). 
%Since the majority of particles leaving Sagittarius A* deposit their energy within $\sim 10$ 
%pc of their source (comfortably within HESS's point-spread-function), we infer that the 
%proton power produced by SA must be $\sim 0.6\times 10^{37}\,(0.5/\epsilon)$ ergs s$^{-1}$, 
%allowing for any possible anisotropy $\epsilon$, either in the acceleration site, or in the 
%spatial distribution of particles propagating away from it, or in the medium surrounding 
%Sagittarius A*. 
The spectrum fits the HESS data and is consistent with observations in other lower energy 
bands. The fact that the required proton power is more than 3 times below that predicted from the 
fitting of the 7 mm data ($\sim 6.0\times 10^{37}$ ergs s$^{-1}$) indicates that most of the 
protons diffuse beyond the HESS point-spread-function, which may explain the recently observed 
diffusion gamma-ray emission from the Galactic Center (Aharonian et al. 2006), constituting one 
of the strongest arguments in favor of the SA model for the origin of Sagittarius A*'s spectrum.

\section{Conclusion and Discussion}
\label{dis}

The combination of a reliable intrinsic source size measurement of Sagittarius A* at 7 mm 
with the detection of TeV emission from its vicinity has important implications for the 
acceleration of relativistic particles near the black hole.  We have shown that SA of 
electrons and protons in a magnetic field dominated plasma can account for both the 7 mm 
observations via {\it in situ} electron synchrotron process and the HESS observations through 
the diffusion of accelerated protons into the surrounding ISM. The fact that a medium with 
$\beta_{\rm A}>1$ is required is intriguing in light of recent MHD simulations of black 
hole accretion (Hawley \& Balbus 2002; Hirose et al. 2004).  These numerical calculations 
show that there are always magnetic-field-dominated coronas above the disk and funnels along 
its symmetry axis. These regions may be the electron and proton acceleration sites. 
Speculating further, our results may also be in line with current concepts of a Poynting flux 
dominated outflow, driven either by an accretion torus or by a spinning black hole (Camenzind 
2004; Semenov, Dyadechkin, \& Punsly 2004; Nakamura \& Meier 2004). 

%However, it is possible that the turbulence coherence length is shorter than the size of the 
%acceleration region, i.e., $Rk_{\rm min}$ could be larger than $2\pi$.  Protons may 
%therefore be accelerated in plasmas with $\beta_{\rm A}< 1$. Our constraints to the magnetic field 
%are weakly dependent of $\beta_{\rm A}$, and the requirement of a low gas density is set by the 
%radio luminosity of Sagittarius A*. A shorter turbulence coherence length therefore only results in 
%a lower proton high-energy cutoff for a given $R$.  On the other hand, the 7 mm emission 
%may not be associated directly with a small acceleration site since it would be 
%self-absorbed. A comprehensive investigation of this aspect is warranted.

Although we suggested that most of the protons producing the TeV emission scatter at 
least once within a region corresponding to the HESS point-spread-function, given the 
uncertainties in the observed properties of the ISM, it is still possible that a significant 
fraction of very high energy protons diffuse outward to even larger radii. Then the proton 
spectral index in the acceleration region would need to be smaller than $2.2$ in order to 
explain the HESS photon spectrum since the proton scattering time increases with energy. However, the 
difference between the proton and photon spectral indexes, which depends on the energy-dependent 
diffusion of the former, is expected to be small (Aharonian \& Neronov 2005b), and our main 
conclusions still hold. In this paper we studied the SA of particles interacting with parallel 
propagating waves. This may not be true at the wave injection scale where the turbulence could be 
isotropic. A comprehensive investigation of this aspect will be presented in a separate paper.

We have shown above that particles energized in a magnetized plasma apparently produce most 
of Sagittarius A*'s emission. This may very well be the case for the accretion flow in many
other compact objects. A thorough investigation of the SA model for particles in a magnetic
field dominated outflow will also play an important role in bridging the above mentioned 
theoretical developments with observations. Eventually, studying SA in such compact 
environments may also reveal the nature of large scale relativistic jets, where the coupling 
of plasma turbulence to the energetic particles could be very strong. 

The radiative consequences of relativistic protons in active galactic nuclei and their
implications on the formation of large scale jets have been studied extensively 
during the past few decades (Begelman, Rudak, \& Sikora 1990; Atoyan \& Dermer 2003). With 
the theory of SA developed here, properties of the turbulent plasma responsible for the proton 
acceleration can be inferred. The constraints to the particle acceleration processes would be 
stricter should the acceleration of electrons by the same turbulence be also considered, as is 
the approach we took here in modeling the broadband spectrum of Sagittarius A*.

We have used the spatially integrated results of SA as a first order approximation in the 
previous investigations. However, to model spatially resolved sources, one must take into 
account the source structure. Monte Carlo simulations of SA may be a good approach in such 
cases and this will likely be the next step in the development of our model. For a complete 
treatment of SA, the properties of the waves must also be addressed. MHD simulations provide a 
complementary vehicle for this problem and when combined with the SA theory, can yield a 
global picture of energy dissipation channeled via turbulence. Investigations of this kind 
will be fruitful in providing a better understanding of many nonthermal astrophysical sources.
With little modification, the model can also be applied to other astrophysical systems where 
the plasma is collisionless and strong turbulence presents. The observed emission spectra can 
be used to determine the properties of the plasma and to reveal the underlying physics 
processes.

\acknowledgments

FM thanks Ray Volkas and Roland Crocker for very helpful discussions. SL is supported by the Director's Fellowship at LANL. This 
research was partially supported by NSF grant ATM-0312344, NASA grants NAG5-12111, NAG5 11918-1 (at Stanford), and 
NSF grant AST-0402502 (at Arizona). FM is very grateful to the University of Melbourne for its 
support (through a Miegunyah Fellowship), and MF is supported by the Hauck Foundation through 
Xavier University.

\begin{figure}[htb]
\begin{center}
\includegraphics[height=14cm]{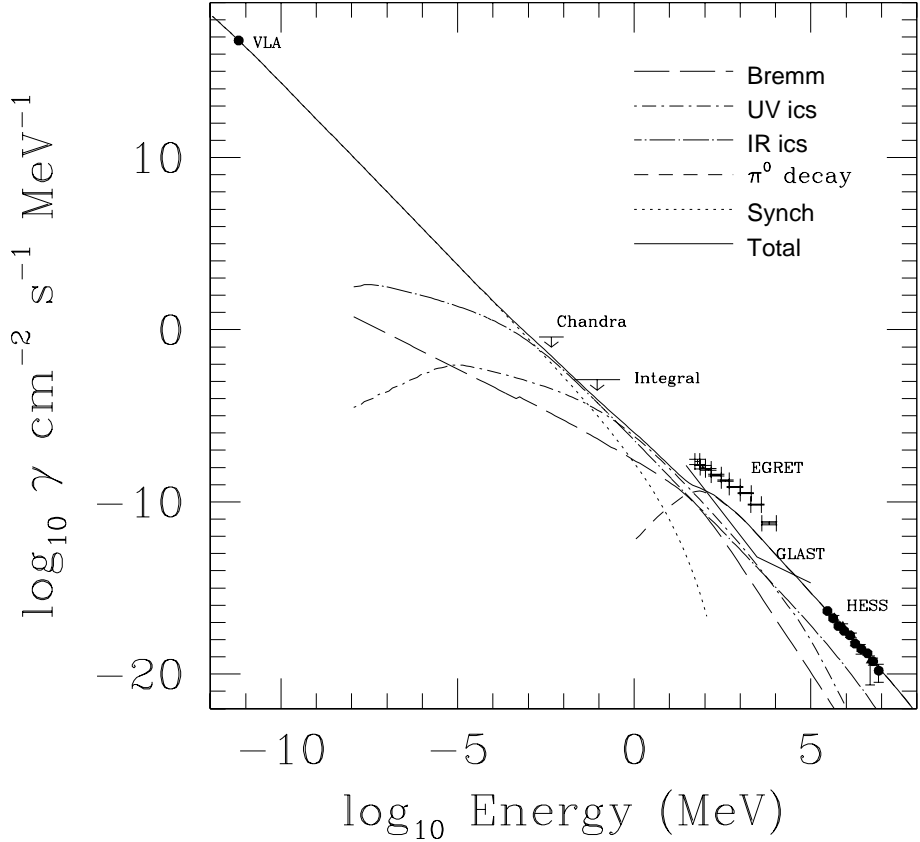}
\end{center}
\caption{Broadband emission from the pp interaction zone due to (short dashed curve) neutral
pion decays, and (long dashed curve) bremsstrahlung, (short dash-dot curve) UV inverse Compton
scattering, (long dash-dot curve) IR inverse Compton scattering, and (dotted curve) synchrotron
emission by the secondary leptons.  The solid curve represents the total broadband emission.   
For comparison, we also show the currently available data (or upper limits), and the expected
GLAST single energy sensitivity for a one-year all-sky survey. The Integral observations   
are from Belanger et al. (2005), the EGRET source 3EG J1746-2852 is from Mayer-Hasselwander et
al. (1998), and the HESS observations are reported in Aharonian et al. (2004). The Chandra
upper limit may be found in Rockefeller et al. (2004), and the VLA datum is reproduced in 
Fatuzzo and Melia (2003).
\label{fig3}
}
\end{figure}

\end{document}